\documentclass[10pt,a4paper,oneside]{book}
\usepackage[utf8]{inputenc}
\usepackage[english]{babel}
\usepackage{amsfonts}
\usepackage{graphicx}

\usepackage[citestyle=authoryear,natbib=true,backend=bibtex]{biblatex}
\bibliography{../../bibliographyAll}

\newcommand{\of}{OpenFOAM\ensuremath{\textsuperscript{\textregistered}}}
\newcommand{\ola}{\textit{olaFlow}}

\graphicspath{{../../figures/}}

\usepackage{graphics}
\usepackage{amssymb}
\usepackage{amsmath}
\usepackage{multirow}
\usepackage{subcaption}

\title{Enhancing active wave absorption in RANS models}
\author{Pablo Higuera}

\begin{document}

\begin{center}
\centering{\Large Enhancing active wave absorption in RANS models}

\par\vspace{7mm}

\large Pablo Higuera

\par\vspace{2mm}

\small \textit{National University of Singapore, Faculty of Engineering\\
Engineering Drive 2, 117578, Singapore.\\
phigueracoastal@gmail.com}
\end{center}

\normalsize

\section*{Abstract}

In this work we review the most common methods for absorbing waves in Reynolds-Averaged Navier-Stokes (RANS) models.
The limitations of active wave absorption, originating from its initial assumption of linear wave theory in shallow waters are overcome and the range of applicability is extended to any relative water depth conditions by re-deriving the formulation.
The new Extended Range Active Wave Absorption (ER-AWA) overperforms the traditional implementation in all the tests performed, which comprise solitary waves and regular waves from shallow to deep waters.
Moreover, the combined use of a relaxation zone and ER-AWA is tested to further reduce wave reflections.
This is most often achieved for a given set of parameters, although some case by case tuning of the relaxation zone parameters would be needed to obtain the best overall performance.
~\\~\\
\textbf{Keywords}

\noindent wave absorption; wave generation; CFD; RANS; OpenFOAM.

\section{Introduction}

Computational fluid dynamics (CFD) modelling has been gaining momentum and attention in recent years as a tool to aid in the design and verification of coastal, marine and offshore structures \citep{atluri09, babei17, dentale18}.
For this purpose numerical models need to be validated so that simulations maintain high fidelity with the challenging wave-driven physics involved.

Since CFD analysis is most often complementary to physical modelling, numerical models usually need to replicate the laboratory setups from experiments.
Furthermore, CFD codes should ideally be able to represent real open-sea conditions at prototype scale as well as a method to evaluate fully realistic conditions and assess the impact of scale effects.
However, this might not be feasible in practice due to computational resources limitations.

Wave absorption is an important feature, required whenever a model performs wave generation.
The generated waves transport energy, which can be reflected at the boundaries if not treated appropriately, increasing the energy level of the system.
Furthermore, the mass imbalance between crests and troughs can produce an increase in the mean water depth for long simulations \citep{mendez01}.
Both phenomena are disadvantageous, especially for long wave simulations, as they will contaminate the data.
Nevertheless, they can be mitigated with the adequate wave absorption techniques.
For example, in laboratory-scale simulations, wave absorption can be implemented in the form of dissipative beaches, passive wave absorbers or active wave absorption systems.
In the case of the open sea, where there are no physical limits, boundary conditions would ideally need to be completely permeable to outgoing waves (open boundary condition).

Some of the most important challenges that CFD faces presently is simulating structures in large water depths ($> 50$ m).
Floating offshore platforms have long been one of the main targets of numerical simulations, where topics as flow-induced motions \citep{kim11} or extreme wave impacts \citep{veldman11} have been studied.
Nevertheless, due to the recent boom of renewable energies, the attention has been widely shifted into floating offshore wind turbines.
These structures are usually moored in deep waters, where where wind energy potential is larger and more predictable, but also where wave conditions are harsher.
Such demanding solicitations required developing new CFD fluid-structure interaction modelling techniques \cite{dunbar15, liuYuanchuan17}. Besides, wave absorption in deep water also poses specific challenges that will be reviewed in depth in the next section.
In short, the effectiveness of active wave absorption methods decreases or requires complex digital filters that need to be optimized for specific wave period ranges.
On the contrary, passive wave absorption remains effective and straightforward to apply.
However, given that wavelengths are longer in deeper waters and this method requires extending the domain at least 1 or 2 wavelengths to obtain an acceptable performance \citep{wei95}, the final computational cost is often excessive.

This paper is structured as follows.
The state of the art of wave absorption in numerical models is reviewed in the next section.
The development of a new active wave absorption (AWA) model with higher performance in deeper waters is explained in the following section.
Afterwards, the numerical model \ola{}, developed within the \of{} framework, will be described.
In Section~\ref{sec:experimentsERAWA} numerical experiments will be analysed to test the performance of the new extended range active wave absorption (ER-AWA).
A coupled system using ER-AWA and relaxation zones as an approach to decrease reflections further is described next.
Finally, conclusions are drawn.

\section{Literature review}

There are two main approaches to absorb waves in numerical models: passive and active absorption.
A comprehensive literature review of wave absorption techniques applied in Reynolds Averaged Navier-Stokes (RANS) numerical models is presented as follows.

Passive wave absorption (PWA) is a classical approach comprised by different techniques.
The simplest approach consists in creating a dissipative beach in which waves will break.
However, longer waves will reflect instead of breaking, transferring energy back into the testing domain.
This approach is widely used in physical tanks, in which space constraints often exist.
Due to these restrictions, CFD models have been used to design and optimize the dissipative beach profiles and compositions \citep{magee15} to enhance the absorption performance.

The second approach is sponge layers or damping zones \citep{israeli81, larsen83}.
In this approach, momentum damping terms acting only in these zones are added to the Navier-Stokes equations.
Normally, the damping rate is controlled by blending functions, which vary smoothly to prevent reflections from discontinuities.
Sponge layers have been traditionally linked with internal wave generation, e.g. \cite{ha13}, in which waves are also generated with localized mass or momentum source terms.
Therefore, waves are radiated in all directions (opposite directions in 2D cases), and require wave absorption at all boundaries.
Sponge layers present several disadvantages.
For example, this technique is known to produce an increase in the mean water level as the simulation progresses \citep{mendez01}.
This can be solved by modifying the continuity equation to include mass sink terms.
Also, effective absorption of longer waves requires longer damping zones, which in turn may increase the computational cost because of the large computational domain.

The final approach is relaxation zones \citep{fuhrman06, jacobsen12}.
In this method, the computational solution to the momentum and Navier-Stokes equations is blended (or relaxed) with a known flow solution in a region near the boundary.
This might be the theoretical expression for particle velocities and free surface elevation given by any wave theory (for wave generation and absorption), or quiescent velocity and fixed still water level (for pure wave absorption).
Different weighting functions and expressions are available in literature, although they all vary smoothly to prevent spurious reflections.
Similarly to sponge layers, relaxation zones present an excellent performance for short wave absorption, but they require to be longer to handle long waves, thus increasing significantly the computational cost of the simulations.

Active wave absorption (AWA) is another technique which poses significant advantages with respect to PWA.
The rationale behind AWA is performing measurements in the flow from which the incident and reflected wave signals can be separated, usually via applying digital filters.
Then, this information (so-called feedback) is used to correct the wave generation boundary condition, in order to absorb the incoming waves while still generating the target waves.

AWA was first applied to RANS in \cite{troch99}, based on the AWASYS system \citep{frigaard94}.
Their method involved measuring velocities inside the wave flume, transforming them into incident and reflected components via a finite impulse response (FIR) filter, and correcting the piston-type wavemaker movement to generate a wave that would cancel out the incoming reflected waves.
The main advantage of AWA is that since it is acting on the boundary, the domain of interest does not need to be extended, thus not increasing significantly the computational costs.

Arguably the simplest AWA technique was presented in \cite{schaffer00}.
Since this method is based on linear wave theory in shallow waters, it will be called shallow water AWA (SW-AWA) in this work.
The system is based on a very simple digital filter, which requires the free surface elevation at the wave generation boundary as its only input.
Despite this method being conceived for piston-type wavemakers in experimental facilities, it has been extensively applied in RANS models afterwards \citep{torres10, didier12, higuera13b, higuera15, miquel18}, both for piston-type moving boundaries and Dirichlet-type static boundary conditions.
Limitations to the practical applicability of AWA, derived from the initial assumption of shallow waters, are well known.
This means that while excellent performance can be achieved when absorbing longer waves, this method is significantly less effective for shorter waves (i.e. deep water conditions).

Additionally, AWA can also be based on other hydrodynamic magnitudes.
\cite{spinneken14} presented a force-feedback method based on an infinite impulse response (IIR) filter which requires the force exerted by the fluid on the wavemaker as input.
One of the strong points of this method is that since force is obtained by integrating pressure, it is less sensitive to local disturbances.
Nevertheless, \cite{spinneken14} also report limitations to generate and absorb highly-nonlinear waves.

Finally, there is a totally different approach for AWA, based on Sommerfeld radiation condition \citep{sommerfeld94}.
This technique, widely applied in Boussinesq-type models, is applied to a RANS model in \cite{luppes10, wellens12}.
One of the advances of their developments is that wave celerity is approximated using digital filters, with which the performance of absorption can be optimized over the ranges of wave numbers of interest.

As a summary, AWA does not increase the computational cost significantly, because it works at the boundaries, while PWA requires enlarging the computational domain.
In this sense, \cite{vyzikas18} found that AWA reduced the computational cost by at least 30\% with respect to relaxation zones.
Moreover, SW-AWA is more effective in absorbing long waves than short waves, while PWA is significantly more effective for short waves.
Finally, it is also worth mentioning that AWA can not absorb wave components that propagate perpendicularly to the boundary, and if they are not treated adequately, they might cause spurious wave generation instead.
A 3D theory to mitigate this effect was presented in \cite{higuera13a}.
Conversely, PWA can dissipate waves in any direction.

Given the benefits and limitations of SW-AWA, the main goal of this work is extending its range of application into deeper water conditions.
Furthermore, in view of the complementary nature of PWA and AWA, the second objective is to explore combinations of both techniques to obtain low reflection coefficients for all relative water depth regimes, while maintaining a reduced computational cost.

\section{Extended Range Active Wave Absorption (ER-AWA)}

The present implementation of SW-AWA in \ola{}, as described in \cite{higuera13a}, is derived from \cite{schaffer00}, which means that it is based on linear wave theory in shallow waters.
This absorption mechanism requires measuring the free surface elevation (FSE) at the boundary and can work coupled simultaneously with wave generation boundary conditions.
The velocity correction equations is:

\begin{equation}
	\label{eq:absOld}
	\Delta U = - \frac{c}{h} \, \Delta\eta = - \sqrt{\frac{g}{h}} \, \Delta\eta
\end{equation}

\noindent where $\Delta U$ is the velocity correction, applied as a constant throughout the water depth, $h$ is the water depth, $c$ is wave celerity, which is $c = \sqrt{g \, h}$ for shallow water conditions, being $g$ the acceleration due to gravity.
$\Delta\eta = \eta_M - \eta_T$ is the difference in wave height due to the unexpected waves incident to the boundary (also called reflected waves), equal to the measured free surface elevation (FSE, $\eta_M$) minus the target FSE ($\eta_T$), corresponding to the target wave generation at the boundary.
For a purely absorbing boundary in which no waves are being generated, $\eta_T = 0$.

As reported in \cite{higuera13a}, reflections outside the shallow water regime into the intermediate waters are moderate, on the order of 10\%.
However, cases closer to deep water conditions but still in intermediate waters can yield reflection coefficients significantly larger than 20\%, which will contaminate the results.
The computational efficiency of AWA is the main driver to seek enhancements for the formulation to be applicable in deep water conditions.

Upon close inspection to equation~\ref{eq:absOld}, two limitations can be detected for deeper water applications.
First, the wave celerity $c = \sqrt{g \, h}$ assumes the non-dispersive wave regime, which is a very convenient simplification, as celerity only depends on the local water depth.
In general, wave celerity can be formulated in terms of the wavelength ($L$) and wave period ($T$): $c = \frac{L}{T}$.
This poses an additional challenge, since the wavelength needs to be obtained from either solving the dispersion relation iteratively:

\begin{equation}
	\label{eq:dispersionRel}
	L = \frac{g \, T^2}{2 \pi} \tanh \left( \frac{2 \pi h}{L} \right)
\end{equation}

\noindent or using any of the explicit approximations available, e.g. \cite{guo02}.
This also means that a new digital filter would need an additional input parameter, the wave period.

The second limitation is that the velocity correction is applied as a uniform velocity throughout the entire water column.
This assumption is reasonable in shallow waters, but extremely inaccurate in deep waters.

\begin{table}
 \begin{center}
  \begin{tabular}{cccc}
    \multirow{2}{*}{} & \multicolumn{3}{c}{Relative water depth} \\
     & Shallow waters & Intermediate waters & Deep waters \\ \hline
    \multirow{2}{*}{Conditions} & $k h < \frac{\pi}{10}$ & $\frac{\pi}{10} < k h < \pi$ & $k h > \pi$ \\
     & $\frac{h}{L} < \frac{1}{20}$ & $\frac{1}{20} < \frac{h}{L} < \frac{1}{2}$ & $\frac{h}{L} < \frac{1}{2}$ \\
    $U/U_0$ & $\frac{1}{kh}$ & $\frac{\cosh[k(h + z)]}{\sinh[kh]}$ & $e^{kz}$ \\
    $W/W_0$ & $1 + \frac{z}{h}$ & $\frac{\sinh[k(h + z)]}{\sinh[kh]}$ & $e^{kz}$ \\
    Sketch (U)& \includegraphics[width=0.2\textwidth]{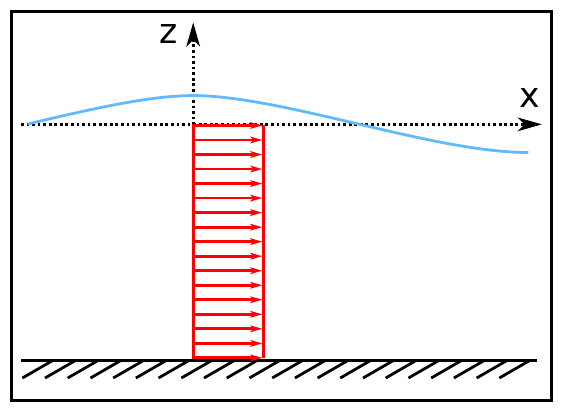} & \includegraphics[width=0.2\textwidth]{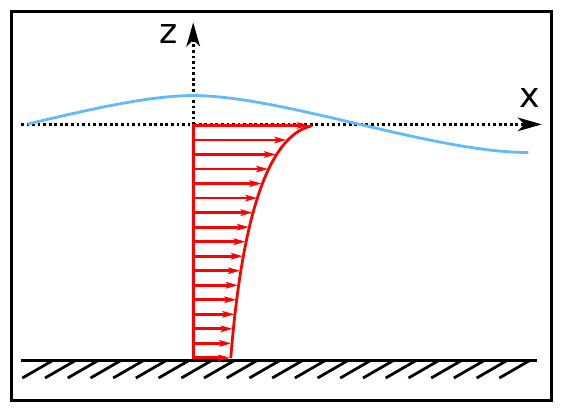} & \includegraphics[width=0.2\textwidth]{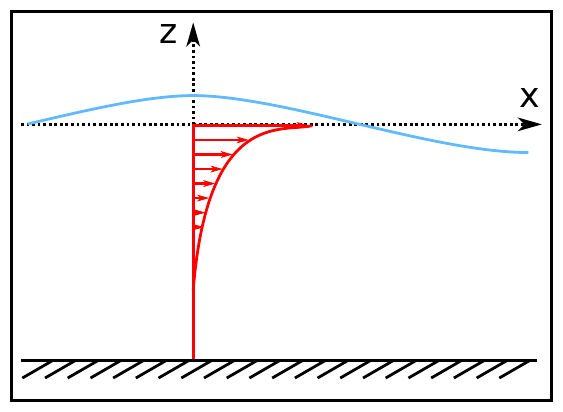} \\
  \end{tabular}
  \caption{Horizontal and vertical velocities at different relative water depths. Sketches are not to scale.}{\label{tab:relativeWaterDepths}}
 \end{center}
\end{table}

In Table~\ref{tab:relativeWaterDepths}, the expressions for the horizontal ($U$) and vertical ($W$) velocity components of Stokes I theory are presented.
The general expression is given for intermediate waters, while the formulas in shallow and deep waters are the simplifications applicable to either conditions.
In these formulas, waves propagate in the positive $x$ direction and the $z$ axis points vertically upwards.
The origin of the coordinate system is located at the still water level, therefore the seabed is at $z = -h$.
Both velocity components have been made dimensionless dividing by $U_0 = \frac{H}{2} \omega \cos[\Phi]$ and $W_0 = \frac{H}{2} \omega \sin[\Phi]$.
Here, $\Phi$ is the phase of the wave, equal to $kx-\omega t + \phi$, where $\phi$ is a phase shift.
As can be observed in the sketches (not to scale), the velocity profile in shallow waters is uniform along the water column, whereas it shows an exponential decay shape in deep waters.
\cite{svendsen06} points out that the horizontal velocity component in deep waters at a water depth $z = -\frac{L}{2}$ is just 4\% of $U_0$.
Consequently, it is conceptually wrong to apply a uniform velocity correction to absorb such waves.

The new implementation presented in this paper, called extended range active wave absorption (ER-AWA), aims at solving the two aforementioned problems.

The first issue is straightforward to resolve: celerity is calculated in the most general way, solving the wavelength with equation~\ref{eq:dispersionRel} and applying $c = \frac{L}{T}$.
The only additional piece of information needed by the new technique is wave period.
Generally, the wave period in numerical simulations of regular waves is given.
If this is not the case, one could start with an estimated wave period and the boundary condition could adjust itself based on the feedback it receives (i.e. the time series of the reflected FSE).
For irregular waves, selecting the peak spectral period is usually advised.
In order to evaluate the variation in performance if the selected wave period is not accurately estimated, a sensitivity analysis to wave period deviations has been performed in Section~\ref{sec:sensitivityT}.

Regarding the second issue, the profile of the velocity correction along the water column, it can also be resolved in an intuitively way.
\cite{schaffer00} provide a simple explanation on the principle behind AWA: waves incident to a vertical wall that produces perfect reflection generate a standing wave, formed by the incident and reflected wave trains, with equal characteristics ($H$, $T$) but travelling in opposite directions.
In order to prevent reflection, the AWA system must cancel out the reflected wave train.
This can be achieved by generating an additional wave train at the boundary, with the same characteristics and still travelling away from the boundary, but with opposite phase.
Therefore, the simple AWA filter in equation~\ref{eq:absOld} is generating in practice  a (shallow water) wave, opposite in phase to the incident waves, thus the negative sign.
Therefore, considering the general linear wave theory horizontal velocity profile in shallow waters:

\begin{equation}
	\label{eq:absProcess1}
	- \frac{c}{h} \, \Delta\eta = \frac{H}{2} \omega \frac{1}{kh} \cos[\Phi].
\end{equation}

By simple manipulations and leaving wave celerity in its general form, the expression can be presented in the following way:

\begin{equation}
	\label{eq:absProcess2}
	- \frac{c}{h} \, \Delta\eta = \frac{c}{h} \frac{H}{2} \cos[\Phi].
\end{equation}

The information about the wave incident to the boundary, represented by $\Delta\eta$, is very limited because it comes from a single-point measurement.
This means that the phase of the incident wave ($\Phi_i$) is unknown, e.g. if $\Delta\eta = l$, where $l$ is any number, the incident wave may have $H = 2l$ and $\Phi _i= 0$ at that instant, or $H = 4l$ and $\Phi_i = \frac{\pi}{3}$, or generally any other combination that fulfils $H = 2l \cos[\Phi_i]$.

Since it is not straightforward to obtain wave phase from a single measurement, although in principle it could be estimated a posteriori analysing dynamically the time series of $\Delta\eta$, an additional assumption is needed in equation~\ref{eq:absProcess2}.
The simplification suggested in this work is to assume that the velocity correction corresponds to the horizontal velocity profile under the wave crest ($\Phi = 0$) for a wave of $H = -2 \Delta\eta$.
Therefore, the new active wave absorption expression, generalized to any water depth is as follows:

\begin{equation}
	\label{eq:ER-AWA}
	\Delta U = - \Delta\eta \, \omega \frac{\cosh[k(h + z)]}{\sinh[kh]},
\end{equation}

\noindent which, as already mentioned, only requires the wave period as an additional input data.

% Integration over a wave period
% This simplification does not produce significant differences in the particle velocity field.
% For example, the maximum punctual and instantaneous difference in particle velocities  between the simplification and using the real phase is  in the order of $10^{-14}$,  throughout the whole range of relative water depths.
 %Using the old formulation yields very similar maximum differences for shallow water conditions, however, differences may grow up to 90\% for deep waters.

\section{Numerical model}

The numerical model used in this work is \ola{} \citep{olaFlow}, conceived as a continuation of the developments in \cite{higueraphd}.
\ola{} is a finite volume Reynolds-Averaged Navier-Stokes (RANS) solver for two incompressible phases developed within the \of{} \citep[]{weller98} framework.
The interface between phases is captured via the Volume of Fluid (VOF) technique \citep{berberovic09}, in which the system is treated as a mixture of both fluids using an indicator function ($\alpha$) marking the content of each cell.
$\alpha=1$ denotes a pure water cell, $\alpha=0$ marks a pure air cell, and $0 < \alpha <1$ represents the interfacial cells.

The continuity, Navier-Stokes and VOF equations solved by the model are as follows:

\begin{equation}
	\label{eq:contOF}
	\nabla \cdot ( \rho \, \bf{U}) = 0,
\end{equation}

\begin{equation}
	\label{eq:momConsOF}
	\begin{split}
		\frac{\partial \rho \bf{U}}{\partial t} 
		& + \nabla \cdot (\rho \bf{UU}) = \\
		& - \nabla \it{p}^{*} 
		- \bf{g} \cdot \bf{r} \, \nabla \rho 
		+ \nabla \cdot (\mu_{\mbox{\tiny eff}} \nabla \bf{U}) 
		+ \sigma \kappa \bf{\nabla \alpha},
	\end{split}
\end{equation}

\begin{eqnarray}
	\label{eq:vofOF}
		\frac{\partial \alpha}{\partial t} 
		+ \nabla \cdot (\alpha \bf{U})
		+ \nabla \cdot [\alpha (1-\alpha) \bf{U}_c]
		= 0,
\end{eqnarray}

\noindent where $\rho$ is the fluid density, calculated as a weighted average of the densities of water and air ($\rho_w$ and $\rho_a$, respectively): $\rho = \alpha \rho_w + (1 - \alpha) \rho_a$.
$\bf{U}$ is the velocity vector, $t$ is time and $\nabla$ denotes the gradient operator.
$p^{*}$ is the pressure in excess of the hydrostatic, namely $p^{*} = p - \rho \, \bf{g} \cdot \bf{r}$, in which $p$ is total pressure, $\bf{g}$ is the gravity acceleration vector and $\bf{r}$ is the Cartesian position vector.
$\mu_{\mbox{\tiny eff}}$ represents the effective dynamic viscosity of the fluid, comprised by the molecular viscosity and the turbulent viscosity given by RANS turbulence models.
The last term in equation~\ref{eq:momConsOF} is the surface tension force, in which $\sigma$ is the surface tension coefficient and $\kappa$ is the curvature of the free surface.
The only new term in equation~\ref{eq:vofOF} is $\bf{U}_c$, a compression velocity aimed at maintaining a sharp interface which acts only at the interface between the fluids ($0 < \alpha < 1$), in the normal direction to the free surface.

The \ola{} model features a library to generate waves as Dirichlet-type boundary conditions for all the theories included in \cite{mehaute76}, extended from \cite{higuera13a, higuera13b}, in which it was initially developed and validated.
Wave generation replicating physical setups such as piston and flap wavemakers was introduced later in \cite{higuera15}.
For both wave generation mechanisms, waves can be absorbed using AWA.
Nevertheless, as discussed in the previous section, the present implementation of SW-AWA inherits severe limitations to be applied in deep water conditions.

\section{ER-AWA performance modelling}
\label{sec:experimentsERAWA}

In this section the performance of the ER-AWA will be tested and compared with previous implementations.
First, the benchmark cases for regular and solitary waves in \cite{higuera13a} will be replicated.
The aim of these initial simulations is to characterize the difference in performance between the present implementations of SW-AWA and ER-AWA in \ola{} and the older version of SW-AWA used in \cite{higuera13a}, from which \ola{} derives.
Next, a sensitivity analysis regarding the input parameter for ER-AWA will be performed.
Finally, new regular wave experiments will be tested to explore the performance of SW-AWA and ER-AWA in deep water conditions.

\subsection{Solitary waves benchmark cases}
\label{sec:experimentsSolitary}

In this first section the solitary wave experiments in \cite{higuera13a} will be simulated again, for the present implementation of SW-AWA and the new ER-AWA.

The mesh, setup and numerical parameters used in the benchmark cases have been tried to keep identical to those in \cite{higuera13a}.
The two-dimensional mesh, which is 20.62 m long and 0.70 m high, is formed by hexahedral cells of 2 cm x 1 cm, and totals 70,000 cells.
Regarding boundary conditions, waves are generated at the leftmost ($X = 0$ m) boundary and absorbed at the opposite end of the flume, the bottom is no-slip and the top boundary has an atmospheric (zero pressure fixed value) boundary condition.

Two solitary waves have been tested for a fixed water depth of $h = 0.40$ m, with wave heights equal to $H =$ 0.05 m and 0.15 m.
Although \ola{} includes third order solitary wave generation, the first order Boussinesq theory \citep{boussinesq} has been applied for the sake of consistency with the results in \cite{higuera13a}.
The incident-reflected analysis has been performed at a wave gauge located at $X = 7.5$ m, calculating the ratio between the highest reflected amplitude and the incident solitary wave height.

The original SW-AWA results in \cite{higuera13a} reported a reflection coefficient of 1.51\% for the $H = 0.05$ m solitary wave and 2.63\% for $H = 0.15$ m.
As an initial test, the solitary wave tests have been re-run with the most recent version of SW-AWA in \ola{}, to analyse any differences that may arise from the code changes introduced in the wave generation and absorption boundary conditions during \ola{} development, and those resulting from code changes in the \of{} base implementation.
The present \ola{} SW-AWA version yields reflection coefficients equal to 1.69\% and 2.51\% for the $H = 0.05$ m and $0.15$ m cases.
This points out a very similar performance with respect with the implementation in \cite{higuera13a}.

The new ER-AWA system simulations require a wave period as input.
In theory, the wave period and wavelength of a solitary wave is infinite, however, an effective period and wavelength can be defined given some criteria.
For example, in \cite{deandalrymple}, the effective solitary wavelength is defined as the distance at which 95\% of the volume of the infinitely long wave is contained, namely $L = 4.24 \frac{h}{\sqrt{H/h}}$, which would yield $L = 4.80$ m and 2.77 m for the $H = 0.05$ m and $0.15$ m cases, respectively.
The criterion used in \ola{} is $L = \frac{4 \pi}{\sqrt{3}} \frac{h}{\sqrt{H/h}}$, i.e. the initial elevation of free surface is 1.5\% of $H$, which yields $L = 8.21$ m and 4.74 m instead.
The equivalent wave periods can be calculated dividing $L$ by the wave celerity, $c = \sqrt{g (h + H)}$: 3.91 s and 2.04 s.
Using these values as input, the ER-AWA system is able to reduce the reflection coefficients down to 1.41\% and 1.50\% for the $H = 0.05$ m and $0.15$ m solitary waves.

\subsection{Regular waves benchmark cases}

\begin{figure}
  \center
  \includegraphics[width=0.9\textwidth]{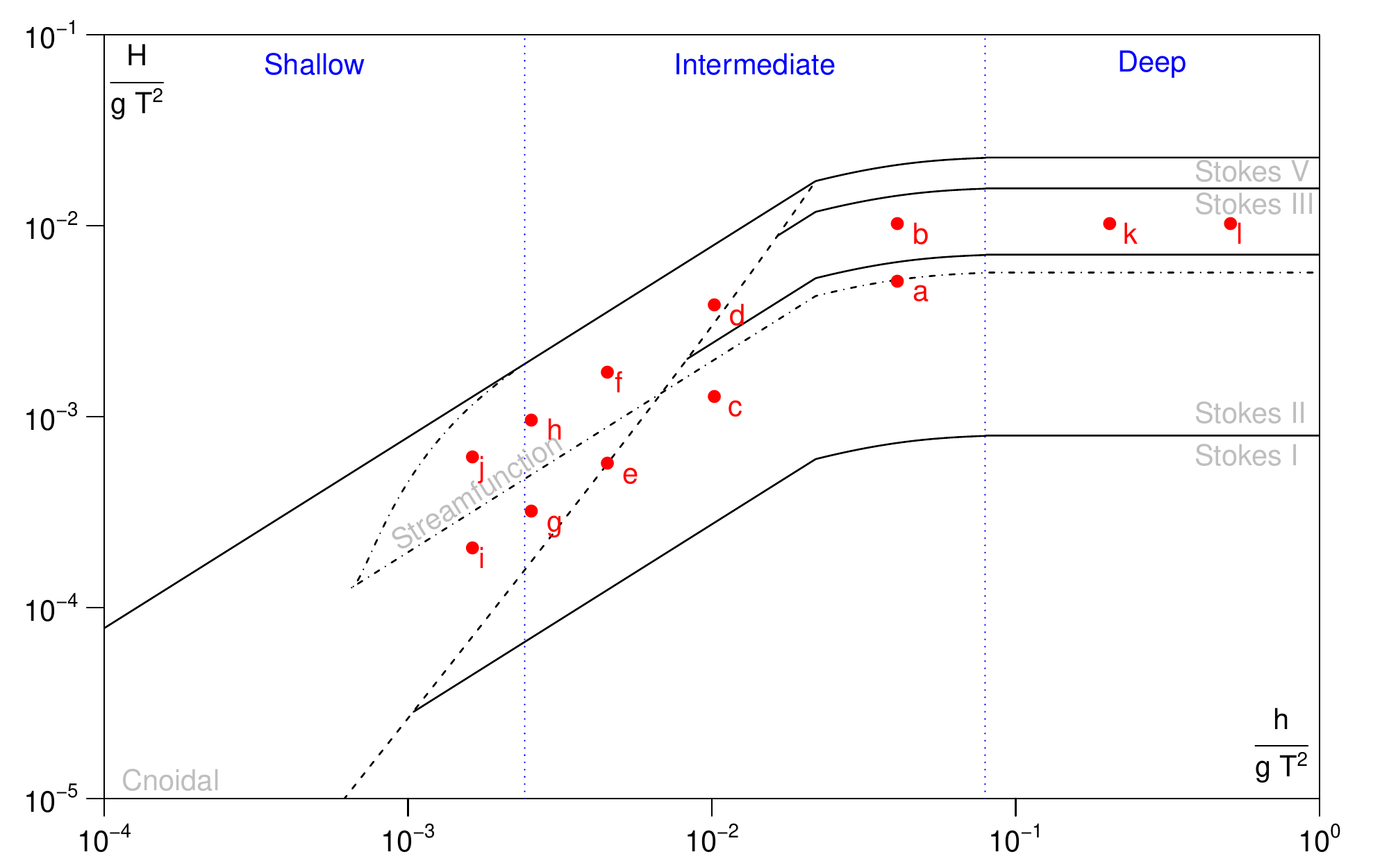}
  \caption{\label{fig:waveTheoryFig} Wave theory for each case tested, according to \cite{mehaute76} classification. Cases referenced as [$H$ (m), $T$ (s), $h$ (m)]: (a) [0.05, 1, 0.4]; (b) [0.10, 1, 0.4]; (c) [0.05, 2, 0.4]; (d) [0.15, 2, 0.4]; (e) [0.05, 3, 0.4]; (f) [0.15, 3, 0.4]; (g) [0.05, 4, 0.4]; (h) [0.15, 4, 0.4]; (i) [0.05, 5, 0.4]; (j) [0.15, 5, 0.4]; (k) [0.10, 1, 2]; (l) [0.10, 1, 5].}
\end{figure}

The regular wave conditions tested are the combination of two wave heights, $H =$ 0.05 m and 0.15 m, and five wave periods, $T =$ 1 s, 2 s, 3 s, 4 s and 5 s, for a water depth of $h = 0.40$ m.
From here on, specific cases will be represented by its wave parameters: [$H$ (m), $T$ (s), $h$ (m)].
The case $T =$ 1 s and $H = 0.15$ m is close to the breaking limit and the incident-reflected wave interaction produced significant wave breaking, therefore, this test has been substituted by $H = 0.10$ m and its results are indicated with an asterisk ($^*$) hereinafter.
All these conditions correspond to intermediate and deep waters, as indicated in figure~\ref{fig:waveTheoryFig}.

Free surface elevation has been sampled at three gauges to perform the incident-reflected analysis.
The gauge locations, which are case-dependent due to the changing wavelength ($L$) as a function of wave period, have been calculated accordingly to \cite{mansard80} method.
The first gauge is always located at $X = 7.5$ m; the second and third gauges are positioned at a distance $L/10$ and $L/4$ from the first one.
No turbulence model has been employed, since the flow conditions have been deemed laminar.
The simulations are run for at least 30 wave periods with a Courant number of 0.3, and take approximately 10 hours each in serial (Xeon processor, 2.50 GHz).

\begin{table}
    \begin{subtable}[b]{0.40\textwidth}
        \centering
          \begin{tabular}{llcc}
                & & \multicolumn{2}{c}{H (m)} \\
                & & 0.05 & 0.15 \\ \cline{3-4}
               \multirow{5}{*}{T (s)} & 1 $\quad$ & - & - \\
                & 2 $\quad$ & 4.6\% & 11.2\% \\
                & 3 $\quad$ & 3.8\% & 7.3\% \\
                & 4 $\quad$ & 2.3\% & 6.7\% \\
                & 5 $\quad$ & - & -
         \end{tabular}
    \caption{}{\label{tab:KrCoeffPrev}}
    \end{subtable}
    ~
    \begin{subtable}[b]{0.25\textwidth}
        \centering
        \begin{tabular}{cc}
                \multicolumn{2}{c}{H (m)} \\
                0.05 & 0.15 \\ \hline
                21.3\% & 19.7\%$^*$ \\
                3.2\% & 13.1\% \\
                3.3\% & 8.4\% \\
                2.1\% & 6.1\% \\
                1.8\% & 8.3\%
         \end{tabular}
        \caption{}{\label{tab:KrCoeffOld}}
    \end{subtable}
    ~
    \begin{subtable}[b]{0.25\textwidth}
        \centering
        \begin{tabular}{cc}
                \multicolumn{2}{c}{H (m)} \\
                0.05 & 0.15 \\ \hline
                11.6\% & 18.6\%$^*$ \\
                2.7\% & 10.1\% \\
                2.9\% & 7.5\% \\
                2.0\% & 5.8\% \\
                1.8\% & 8.3\%
         \end{tabular}
        \caption{}{\label{tab:KrCoeffNew}}
    \end{subtable}
    \caption{Reflection coefficients for \cite{higuera13a} (a), SW-AWA (b) and ER-AWA (c). The values indicated by $^*$ correspond to $H = 0.10$ m, since waves break for $H = 0.15$ m.}{\label{tab:KrCoeff1}}
\end{table}

Table~\ref{tab:KrCoeff1} shows the reflection coefficients obtained for all the initial benchmark cases.
When comparing Subtable~\ref{tab:KrCoeffPrev} and Subtable~\ref{tab:KrCoeffOld}, it is clear that \ola{} SW-AWA implementation presents almost identical performance with respect to SW-AWA in \cite{higuera13a}.
Similarly to what happened in the previous section, some reflection coefficients decrease (all cases with $H = 0.05$ m), while others increase, as case [0.05, 2, 0.4], by almost 2\%.
As mentioned before, these small differences lie in source code changes in \ola{} and \of{}.

Two new sets of cases with $T = 1$ s and 5 s, were also simulated in this work to test conditions closer to deep waters and shallow waters, respectively.
As expected, the SW-AWA implementation (Subtable~\ref{tab:KrCoeffOld}) does not perform well in the vicinity of deep waters, with reflection coefficients larger than 20\%.
Conversely, the absorption is adequate in shallow waters.

The new ER-AWA reflection coefficients, presented in Subtable~\ref{tab:KrCoeffNew}, overperform almost all of the previous results.
ER-AWA works remarkably well for case [0.05, 1, 0.4], in which reflection drops from 21.3\% to 11.6\%,
It is also noteworthy that the performance of SW-AWA and ER-AWA are identical in the shallow water cases ($T = 5$ s).
These results prove that the new theory can be applied in a wider range of relative water depth conditions without compromising the original performance of SW-AWA.

Overall, the largest reflection coefficients are obtained for the largest wave heights.
Moreover, reflection coefficients generally increase for decreasing wave periods.
Both conditions point out an important limitation of the present AWA theories. Since they are derived from linear wave theory the performance decreases significantly for the most nonlinear cases, which deviate significantly from the original assumption.

\begin{figure}
  \center
  \includegraphics[width=0.9\textwidth]{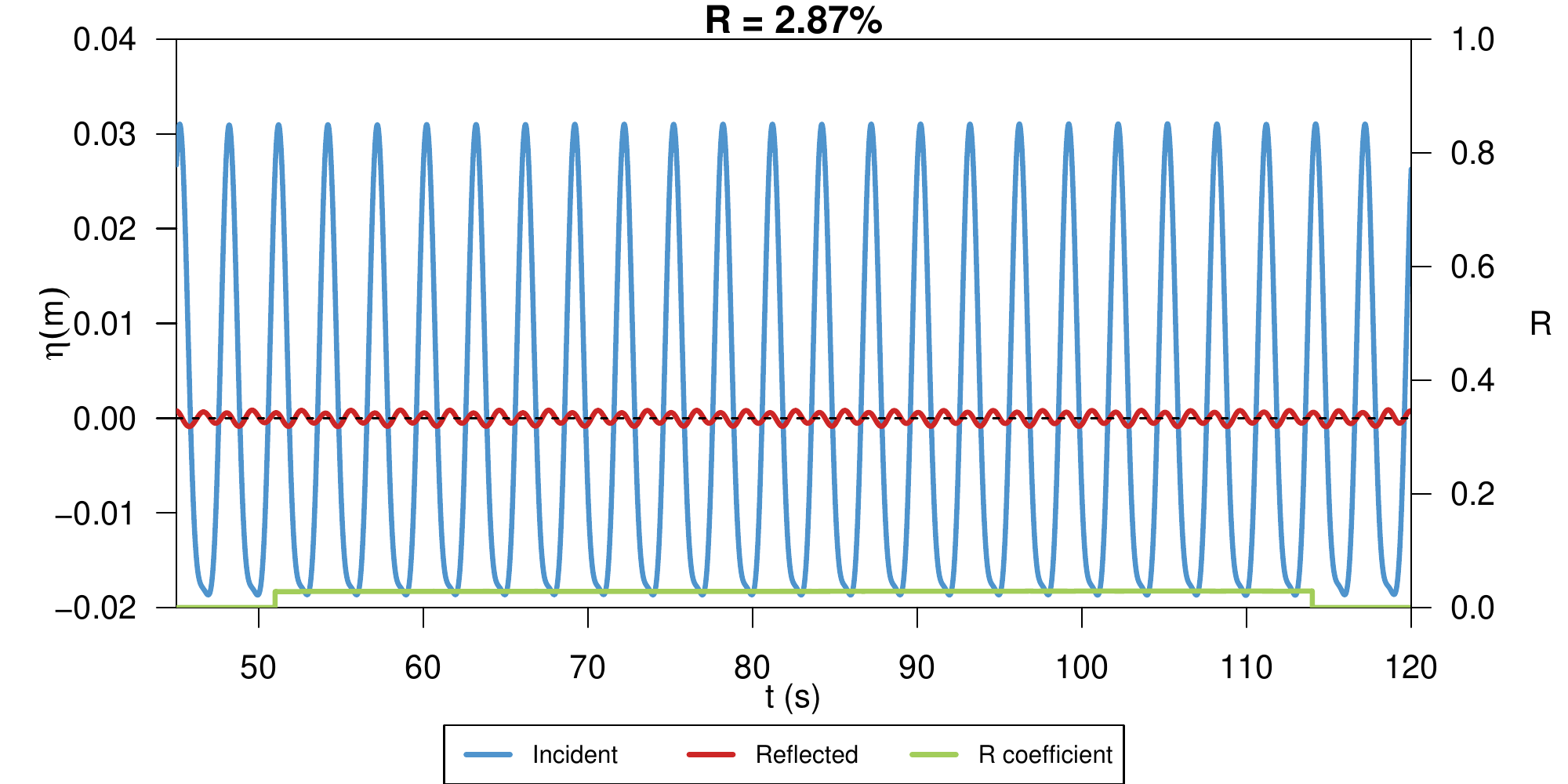}\\
  \includegraphics[width=0.9\textwidth]{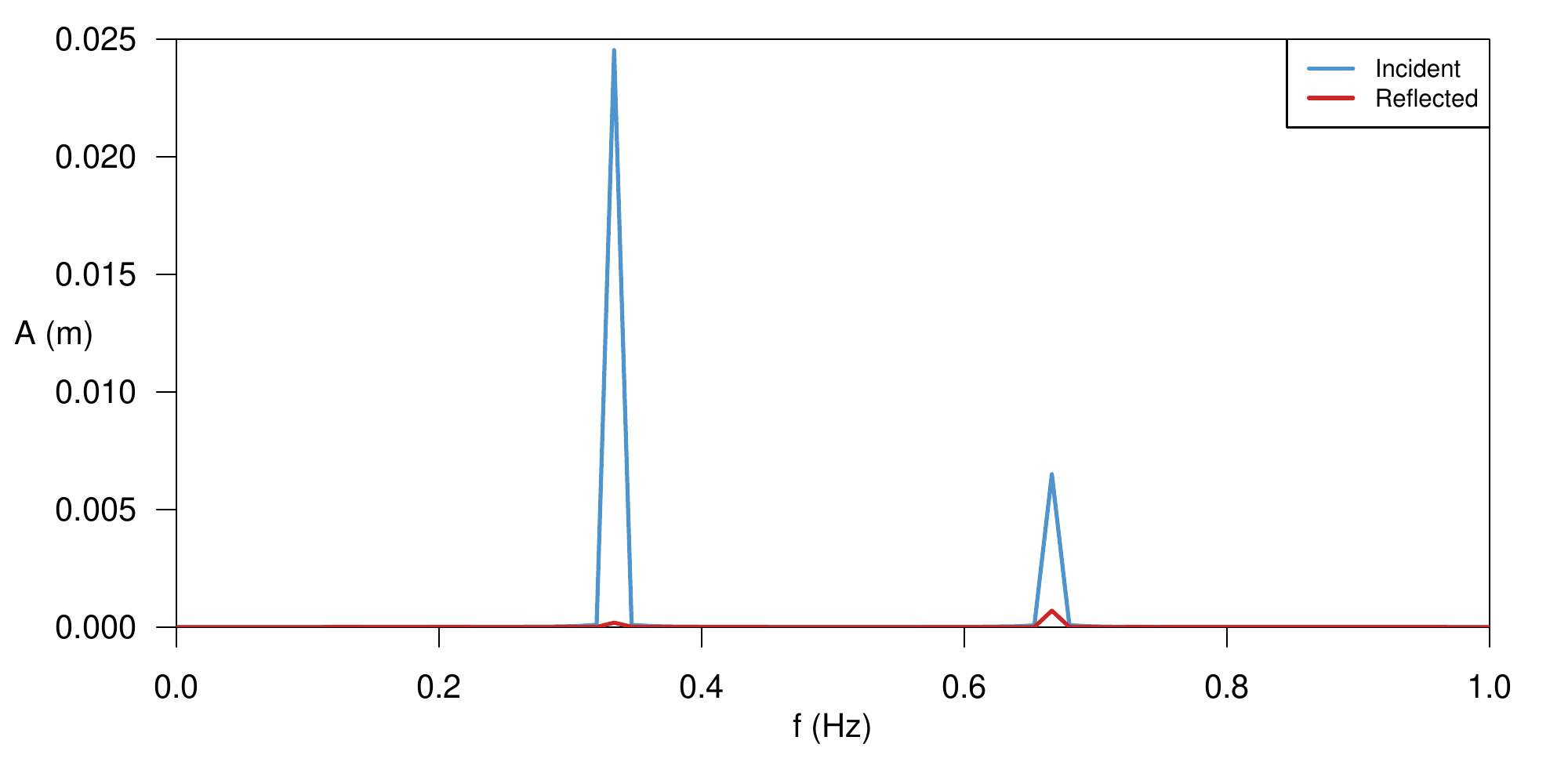}
  \caption{\label{fig:incidentReflectedAnalysis} Incident and reflected analysis for case ER-AWA [0.05, 3, 0.4].
  Top panel: time domain analysis. Bottom panel: amplitude spectral analysis.}
\end{figure}

In Figure~\ref{fig:incidentReflectedAnalysis}, the incident-reflected analysis outputs of case ER-AWA [0.05, 3, 0.4] are presented.
In the top panel, the time domain is shown.
The incident and reflected free surface elevation time series (blue and red lines, respectively) are referenced to the left axis, whereas the time history of the reflection coefficient (green line) is referenced to the right axis.
It can be noted that the quasi-steady state has been reached, as all the incident and reflected waves have a constant amplitude and shape throughout time and that the wave height is very close to the target $H = 0.05$ m.

The bottom panel includes the incident and reflected amplitude spectra.
Two peaks, corresponding to the main harmonic ($f = 1/3$ Hz) and a superharmonic with double the frequency ($f = 2/3$ Hz).
The absorption rate for the main harmonic is almost perfect, while the reflected energy for the second harmonic is noticeably larger in both relative and absolute terms.
This is not always the case, since often the second superharmonic will experience an almost perfect reflection, as later described in Section~\ref{sec:regDeepWater}.
Finally, the reflection coefficient, $R =$ 2.87\%, is calculated following the definition in \cite{goda76}, as the ratio between $H_{m0}$ (four times the square root of the zeroth order moment (i.e. area) of the wave power spectrum) of the reflected spectrum over the incident spectrum.

There are no deep water tests in this initial benchmark due to water depth limitations, which were selected in \cite{higuera13a} within the typical range used in shallow and intermediate water flumes and basins.
Nevertheless, deep water conditions will be explored in Section~\ref{sec:regDeepWater}.

\subsection{Sensitivity analysis to wave period deviations}
\label{sec:sensitivityT}

In this section we will investigate the impact in wave absorption performance of using a wave period that deviates from the real wave generation period as input of ER-AWA.
For this purpose, the same series of benchmark cases from the previous section have been run again, but using $T \pm \Delta T$, with $\Delta T$ equals to 10\% of $T$, as the input period of ER-AWA.

\begin{table}
    \begin{subtable}[b]{0.50\textwidth}
        \centering
          \begin{tabular}{llccc}
                & & \multicolumn{3}{c}{$\Delta T/T$} \\
                & & -10\% & 0 & +10\% \\ \cline{3-5}
               \multirow{5}{*}{T (s)} & 1 $\quad$ & 12.8\% & 11.6\% & 14.0\% \\
                & 2 $\quad$ & 3.4\% & 2.7\% & 2.6\% \\
                & 3 $\quad$ & 2.9\% & 2.9\% & 2.8\% \\
                & 4 $\quad$ & 2.5\% & 2.0\% & 2.2\% \\
                & 5 $\quad$ & 1.6\% & 1.8\% & 2.2\% \\
         \end{tabular}
    \caption{}{\label{tab:KrCoeffSens05}}
    \end{subtable}
    ~
    \begin{subtable}[b]{0.35\textwidth}
        \centering
        \begin{tabular}{ccc}
                \multicolumn{3}{c}{$\Delta T/T$} \\
                -10\% & 0 & +10\% \\ \hline
                18.7\%$^*$ & 18.6\%$^*$ & 18.7\%$^*$ \\
                10.3\% & 10.1\% & 10.7\% \\
                7.3\% & 7.5\% & 7.4\% \\
                5.8\% & 5.8\% & 5.6\% \\
                7.7\% & 8.3\% & 8.6\% \\
         \end{tabular}
        \caption{}{\label{tab:KrCoeffSens15}}
    \end{subtable}
    \caption{Sensitivity analysis of the reflection coefficients for 10\% variations in the input wave period for ER-AWA. (a) corresponds to $H = 0.05$ m and (b) to $H = 0.15$ m, except for the cases indicated by $^*$, which correspond to $H = 0.10$ m.}{\label{tab:KrCoeffSens}}
\end{table}

The results of the sensitivity analysis are shown in Table~\ref{tab:KrCoeffSens}.
The variations, both for the 10\% deficit or excess in wave period, are generally within 10\% of the reflection coefficients measured, and in most cases significantly less than 1\% in absolute value.
The original results, obtained using the target wave period as input, do not show the best performance consistently.
However, the global minima are usually just 0.2\% below, which can be considered as virtually the same value.
The reason is that at that level of precision, on the order of thousandths of the reflection coefficient, there are possibly other uncertainties that might also have an influence on the results.

As a consequence, it can be concluded that small deviations in the input period can be tolerated, since the reflection coefficient is not extremely sensitive to them.

\subsection{Regular waves deep water cases}
\label{sec:regDeepWater}

Two additional cases have been run to test the performance of AWA for deep water cases.
The new cases have larger water depths, as simulating deep water conditions for $h = 0.4$ m required extremely small wave heights and short wave periods.
The new wave conditions (h and l in Figure~\ref{fig:waveTheoryFig}) have $H = 0.10$ m and $T = 1$ s, and different water depth: $h = 2$ m and $h = 5$ m, respectively.
The computational parameters have been kept identical as those of previous cases, including the constant cell size (2 x 1 cm), therefore, the new meshes have a larger number of cells due to the extended vertical domain.

Shallow water absorption was tried first.
For case [$H = 0.10$ m, $T = 1$ s, $h = 2$ m] the reflection coefficient is 44.6\%, while for [$H = 0.10$ m, $T = 1$ s, $h = 5$ m] it reaches 51.8\%.
As expected, the reflection coefficients obtained are extremely large, because the SW-AWA assumptions are far from the conditions tested in terms of relative water depth, thus, neither the wave celerity nor the correction velocity profile are well represented.
Using the ER-AWA formulation diminishes the reflection coefficients notably, to 13.6\% and 13.4\%, proving that the new formulation presents significant advantages.

\begin{figure}
  \center
  \includegraphics[width=0.9\textwidth]{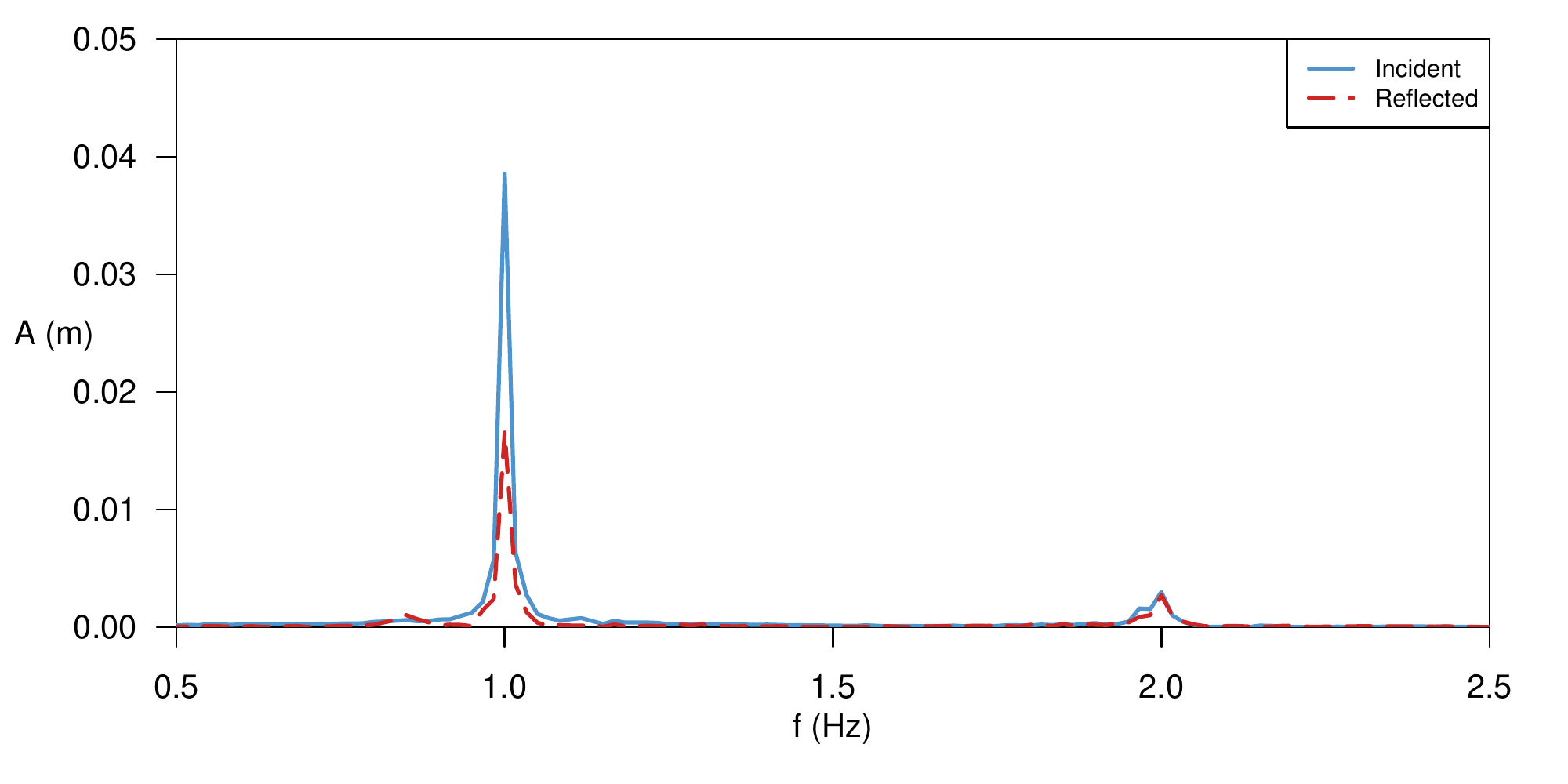}\\
  \includegraphics[width=0.9\textwidth]{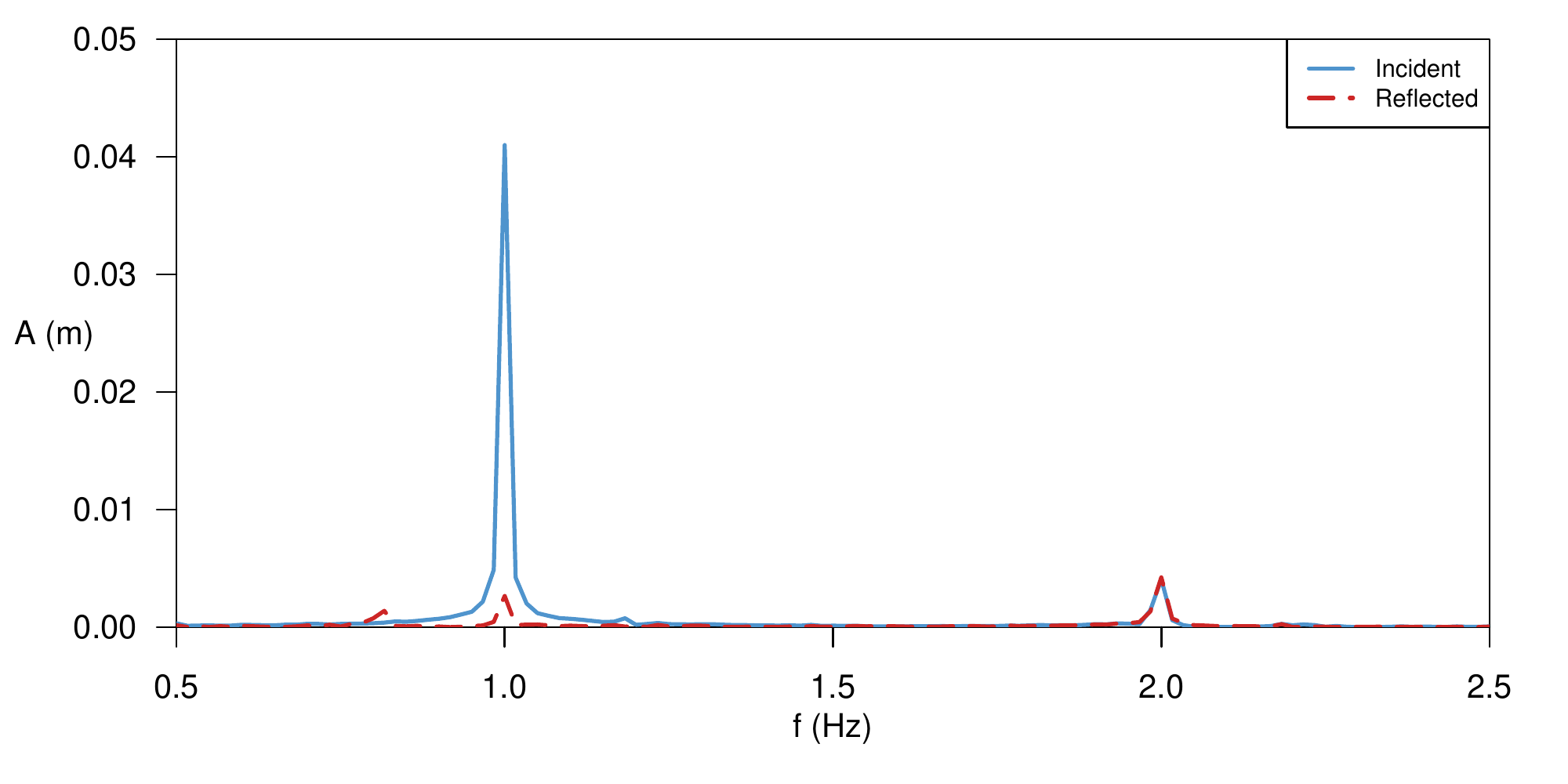}
  \caption{\label{fig:spectrah2} Incident-reflected amplitude spectrum decomposition for case [0.10, 1, 2]. Top panel: SW-AWA. Bottom panel: ER-AWA.}
\end{figure}

The incident and reflected amplitude spectrum decomposition for both AWA techniques in case [0.10, 1, 2] is included in Figure~\ref{fig:spectrah2}.
This monochromatic sea state results in a main harmonic at $f = 1$ Hz and a superharmonic located at $f = 2$ Hz.
Regarding the incident-reflected analysis, both cases share a common feature, only the main harmonic is attenuated by the AWA boundary condition, while the amplitude of the superharmonic is virtually identical in the incident and reflected spectra.
As mentioned before, the global reflection coefficient takes into account all the frequencies.
However, taking into account the main harmonic only, the case [0.10, 1, 2] yields reflection coefficients of 43.1\% and 6.5\% for SW-AWA and ER-AWA, and case [0.10, 1, 5], 38.8\% and 3.8\%.

% According to the amplitudes of each of the components, a 16\% and 9\% deficit in the target incident wave height ($H = 0.10$ m) can be detected for the SW-AWA and ER-AWA cases.
% This could be resolved by fine tuning the incident wave height used as input for the wave generation boundary condition.

\section{AWA combined with relaxation zones}

In view of previous results it is obvious that the new formulation for AWA represents an advantage when simulating waves at any relative water depth conditions.
However, the performance might not always be as good as required for engineering design purposes in case of large wave nonlinearity, as reflection coefficients significantly larger than 10\% have been obtained.
Given that AWA is more effective for long waves, as opposed to PWA, which performs better for shorter waves, a combination of both methods can be proven advantageous, minimizing the length of the relaxation zone required while increasing the performance of AWA alone.
The idea of combining AWA and PWA is not new, as \cite{israeli81, lin99} already pointed out that the best absorption performance in their tests was obtained with the combination of a sponge layer and a radiation boundary condition.

\subsection{Relaxation zone description}

In this work, a very simple relaxation zone has been developed to work for pure wave absorption.
The concept behind relaxation zones is simple: if a theoretical solution for the flow is known within a region, the solution of the Navier-Stokes equations can be gradually blended (or relaxed) with these theoretical values to generate or to absorb waves, as demonstrated in \cite{jacobsen12}.
In the present paper, the target solution within the relaxation zone is a quiescent body of water with the free surface at located the initial still water level.

The relaxation zone technique works as follows.
The VOF function ($\alpha$) and velocity ($u$) are corrected explicitly after the VOF function advection and prior to solving the pressure Poisson equation using the following generic expression:

\begin{equation}
	\label{eq:relaxMethod}
	\Lambda = w_R \, \Lambda_{\tiny\mbox{NS}} + (1 - w_R) \, \Lambda_{\tiny\mbox{TH}},
\end{equation}

\noindent where $\Lambda$ is the variable of interest and the subscripts \textit{NS} and \textit{TH} represent the Navier-Stokes solution and the theoretical values, respectively.
The variable $w_R$ is a weight function that varies smoothly within the relaxation zone.
There are numerous formulations available in literature for $w_R$.
Since the goal of this paper is not selecting the best-performing expression, the only formulation used in this work will be \cite{fuhrman06}:

\begin{equation}
	\label{eq:relaxExpression}
	w_R = 1 - \frac{e^{\sigma^P} - 1}{e - 1}.
\end{equation}

\noindent where, $\sigma$ represents a coordinate system relative to the relaxation zone such that $w_R(\sigma = 0) = 1$ at the interface between the relaxation zone and the domain of interest.
Generally, it is also the case that $w_R(\sigma = 1) = 0$ at the relaxation zone end boundary, although this does not apply to the present work.
The parameter $P$ controls the shape of the relaxation function.
Optimizing the value of $P$ for a given case is also not on the scope of this paper, therefore, the default value recommended and proven to be suitable for wave generation and absorption in \cite{jacobsen12}, $P = 3.5$, will be applied.

There is a main difference between the application of the relaxation zone in \cite{jacobsen12} and in this work.
In \cite{jacobsen12} the end of the relaxation zone corresponds to $\sigma = 1$, therefore, $w_R = 0$ at that location, and the variables $\alpha$ and $u$ take the theoretical value.
This will produce reflections if the dissipation zone is not long enough.
In this work the relaxation zone will only be partially inside the mesh, so that at the end boundary $0 < \sigma < 1$ and $w_R > 0$.
This will produce some energy dissipation, specially on shorter waves and superharmonics, while letting the relaxation zone be permeable to longer waves, which will be absorbed by the ER-AWA at the end boundary.

\subsection{ER-AWA and relaxation zone experiments}

In this section, all the previous tests will be simulated again with a combination of the ER-AWA and a relaxation zone (ER-AWA-RZ).
From preliminary tests it has been observed that a relaxation zone of length equal to the target wavelength ($L$), placed $L/3$ inside the mesh domain, generally enhances the performance of active wave absorption.
Therefore, this setup has been used in all the following simulations.
It must be noted that the target of this section is not to obtain the best absorption performance, but to introduce the benefits of a coupled ER-AWA-RZ system.
Therefore, a fine tuning analysis based on each case conditions and limitations should be performed to obtain the best wave absorption results.

\begin{table}
    \begin{subtable}[b]{0.50\textwidth}
        \centering
          \begin{tabular}{llcc}
                & & ER-AWA & ER-AWA-RZ \\ \cline{3-4}
               \multirow{5}{*}{T (s)} & 1 $\quad$ & 11.6\% & 10.1\% \\
                & 2 $\quad$ & 2.7\% & 4.5\% \\
                & 3 $\quad$ & 2.9\% & 2.8\% \\
                & 4 $\quad$ & 2.0\% & 1.8\% \\
                & 5 $\quad$ & 1.76\% & 1.0\%
         \end{tabular}
    \caption{}{\label{tab:KrCoeffRelax05}}
    \end{subtable}
    ~~~
    \begin{subtable}[b]{0.35\textwidth}
        \centering
        \begin{tabular}{cc}
                ER-AWA & ER-AWA-RZ \\ \hline
                18.6\%$^*$ & 18.7\%$^*$  \\
                10.1\% & 8.1\% \\
                7.5\% & 5.2\% \\
                5.8\% & 4.3\% \\
                8.3\% & 4.3\%
         \end{tabular}
        \caption{}{\label{tab:KrCoeffRelax15}}
    \end{subtable}
    \caption{Reflection coefficients for the ER-AWA and ER-AWA-RZ. (a) corresponds to $H = 0.05$ m and (b) to $H = 0.15$ m, except for the cases indicated by $^*$, which correspond to $H = 0.10$ m.}{\label{tab:KrCoeffRelax}}
\end{table}

The comparison between ER-AWA-RZ and the previously-reported ER-AWA reflection coefficient is included in Table~\ref{tab:KrCoeffRelax}.
Generally, adding the relaxation zone enhances the global absorption rate, except for case [0.05, 2, 0.4] in which the reflection coefficient increases from 2.7\% to 4.5\%.
Two main trends can be observed.
First, the reduction in the reflection coefficient increases with the period, since the wavelength and, thus, the length of the relaxation zone also increase.
Also, the reflection coefficient reduction is larger for the largest wave heights (Table~\ref{tab:KrCoeffRelax15}).
This is a result of such wave conditions being away from the linear wave theory assumptions, from which AWA is derived.
Therefore, the absorption performance increases because relaxation zones are not constrained by such limitations.

Regarding the solitary wave cases, the relaxation zone length is set in terms of the effective wavelength, defined in Section~\ref{sec:experimentsSolitary}.
The absorption performance of ER-AWA was remarkable, with reflections below 1.5\%.
Adding the relaxation zone yields reflection coefficients of 2.88\% for $H =$ 0.05 m and 1.45\% for $H =$ 0.15 m, therefore the results produce a notable increase in wave reflections for $H =$ 0.05 m and a slight decrease for $H =$ 0.15 m.

Finally, with respect to the deep water cases in Section~\ref{sec:regDeepWater}, adding the relaxation zone does not improve the results.
For example, in case [0.10, 1, 2], the reflection coefficient is virtually the same, increasing from 13.6\% to 13.7\%.
In case [0.10, 1, 5], however, the absorption performance decreases, as adding the relaxation zone increases the reflections from 13.4\% to 15.0\%.
Nevertheless, as explained before, short period waves are less effected by the relaxation zone set of parameters selected.

As a summary, coupling ER-AWA with a relaxation zone generally enhances further the wave absorption performance, especially for higher and longer period waves.
However, the performance does not always increase for the tested relaxation zone parameters, therefore, such parameters (i.e., length, location and $P$ factor) or even the relaxation zone base function need to be carefully selected on a case by case basis.

\section{Conclusions}

In this paper we analysed the different wave absorption techniques available for RANS models.
Passive wave absorption is most often straightforward to apply and does not rely on initial assumptions.
However, long extensions of the numerical domain on the order of 2 wavelenghts are required to achieve a high level of absorption.
On the contrary, active wave absorption acts at the boundaries, thus, not increasing the computational cost of the simulations.
The performance of active wave absorption depends heavily in its initial assumptions, which were presently linear wave theory in shallow waters.
Hence, the absorption level is low for deep water conditions.

An extension to active wave absorption has been proposed, to extend its applicability to all relative water depths.
This generalization requires and additional input, the wave period of the waves to absorb.
A sensitivity analysis indicates that small deviations in the input period can be tolerated, as the reflection coefficient is not extremely sensitive to them.

The new ER-AWA formulation has been found advantageous, reducing the reflection coefficients obtained for SW-AWA in all the cases for regular and solitary waves tested.
The increase of performance is specially remarkable in deep water cases, as expected.
Nevertheless, ER-AWA has been derived from linear wave theory, hence, as wave nonlinearity increases the reflection coefficient also increases.

Adding a relaxation zone before the ER-AWA boundary condition is most often advantageous because it helps to further reduce the reflections just requiring enlarging the domain an order of magnitude of a third of the wavelength. 
Nevertheless, this is not always the case for the unique set of parameters tested in this work.
Therefore, performing a sensitivity analysis of the relaxation zone parameters is recommended on a case by case basis to obtain the best results.

In view of the results, adopting ER-AWA is recommended in cases in which a reasonable estimate for the wave period can be obtained.
Including a relaxation zone is also recommended if the reflection coefficient from the ER-AWA is still high enough that contaminates the results.
Future works will focus in testing ER-AWA to simulate irregular wave sea states.
The new implementations presented in this paper will be releases as open source as part of the \cite{olaFlow} suite.

\printbibliography

\end{document}